\documentclass[
reprint,
superscriptaddress,
 amsmath,amssymb,
 aps,
prl
floatfix,
]{revtex4-1}

\usepackage{amsmath}
\usepackage{marvosym}
\usepackage{graphicx,subfigure}         
\usepackage{amsmath} 
\usepackage{amssymb,stmaryrd}   
\usepackage{color}
\usepackage{wrapfig}
\usepackage{multirow}
\usepackage{mathtools}
\usepackage{tikz}  
\usepackage{xspace}
\usetikzlibrary{arrows,shapes,trees,positioning}  
\usepackage{comment} 
\usepackage{bm}
\usepackage{physics}

\usepackage{hyperref}
\usepackage{textcomp}
\hypersetup{
    colorlinks=true,
    linkcolor=blue,
    citecolor=blue,
    urlcolor=blue,
}

\usepackage{graphicx}%

\usepackage{gensymb}

\newcommand{\mycomment}[1]{}

\newcommand{\ISHE}{\textrm{ISHE}}

\begin{document}

\title{Symmetry-based phenomenological model for magnon transport in a multiferroic}

\author{Isaac A. Harris}
\altaffiliation[Authors]{ contributed equally}
\affiliation{Department of Physics, University of California, Berkeley, CA, USA}

\author{Sajid Husain}
\altaffiliation[Authors]{ contributed equally}
\affiliation{Materials Science Division, Lawrence Berkeley National Laboratory, Berkeley, CA, USA}
\affiliation{Department of Materials Science and Engineering, University of California, Berkeley, CA, USA}
\email{shusain@berkeley.gov}

\author{Peter Meisenheimer}
\affiliation{Department of Materials Science and Engineering, University of California, Berkeley, CA, USA}

\author{Maya Ramesh}
\affiliation{Department of Materials Science and Engineering, Cornell University, Ithaca, USA}

\author{Hyeon Woo Park}
\affiliation{
Department of Physics, Korea Advanced Institute of Science and
Technology, Daejon, Korea
}

\author{Lucas Caretta}
\affiliation{School of Engineering, Brown University, Providence, RI, USA}

\author{Darrell Schlom}
\affiliation {Department of Materials Science and Engineering, Cornell University, Ithaca, USA}

\author{Zhi Yao}
\affiliation{Applied Mathematics and Computational Research Division, Lawrence Berkeley National 
Laboratory, Berkeley, CA, 94720, USA}

\author{Lane W. Martin}
\affiliation{Department of Materials Science and Engineering, University of California, Berkeley, CA, USA}
\affiliation{Department of Materials Science and NanoEngineering, Rice University, Houston, TX, USA}
\affiliation{Department of Physics and Astronomy, Rice University, Houston, TX, USA}
\affiliation{Department of Chemistry, Rice University, Houston, TX, USA}
\affiliation{Rice Advanced Materials Institute, Rice University, Houston, TX, USA}

\author{Jorge \'I\~niguez-Gonz\'alez}
\affiliation{
Materials Research and Technology Department, Luxembourg Institute of Science and Technology, Esch-sur-Alzette, Luxembourg
}
\affiliation{
Department of Physics and Materials Science, University of Luxembourg, Belvaux, Luxembourg
}

\author{Se Kwon Kim}
\affiliation{
Department of Physics, Korea Advanced Institute of Science and
Technology, Daejon, Korea
}

\author{Ramamoorthy Ramesh}%
 \email{rramesh@berkeley.edu}
\affiliation{Department of Physics, University of California, Berkeley, CA, USA}
\affiliation{Department of Materials Science and Engineering, University of California, Berkeley, CA, USA}
\affiliation{Materials Science Division, Lawrence Berkeley National Laboratory, Berkeley, CA, USA}
\affiliation{Department of Physics and Astronomy, Rice University, Houston, TX, USA}
\affiliation{Department of Materials Science and NanoEngineering, Rice University, Houston, TX, USA}
\affiliation{Rice Advanced Materials Institute, Rice University, Houston, TX, USA}


\begin{abstract}
    
Magnons – carriers of spin information – can be controlled by electric fields in the multiferroic BiFeO$_3$ (BFO), a milestone that brings magnons closer to application in future devices. The origin of magnon-spin currents in BFO, however, is not fully understood due to BFO’s complicated magnetic texture. In this letter, we present a phenomenological model to elucidate the existence of magnon spin currents in generalized multiferroics by examining the symmetries inherent to their magnetic and polar structures. This model is grounded in experimental data obtained from BFO and its derivatives, which informs the symmetry operations and resultant magnon behavior. By doing so, we address the issue of symmetry-allowed, switchable magnon spin transport in multiferroics, thereby establishing a critical framework for comprehending magnon transport within complex magnetic textures. 
\end{abstract}

\maketitle

Magnons, quanta of spin waves, have become the ideal carriers of information in future spin-based technologies 
\cite{Chumak_MagnonicDataProc_2017,Khitun_MagnonicLogic_2010,Kruglyak_Magnonics_2010,Pirro_AdvCoherMagnon_2021}. Magnons can carry information through insulating antiferromagnets, which are desirable candidate materials for novel magnetic storage technologies \cite{Jungwirth_AntiferrSpintron_2016,Han_CoherAntiferrSpint_2023,Jungwirth_MultDirAntifSpint_2018}. While manipulating the magnetic order parameter of most antiferromagnets is impractical in device applications, the antiferromagnetic order of multiferroic bismuth ferrite, BiFeO$_3$ (BFO), is switchable by an electric field due to the magnetoelectric coupling between ferroelectricity and antiferromagnetism \cite{Bibes_MagnetoelectricMem_2008,haykal2020antiferromagnetic,Heron_BFOswitching_2014,Spaldin_AdvMagnetoelecMultiferr_2019,Rovillain_EfieldSpinWavesBFO_2010,Manipatruni_MESO_2019}. 
Electric field control of magnon transport through BFO has been recently demonstrated \cite{Rovillain_EfieldSpinWavesBFO_2010,Parsonnet_Nonvolatile_2022,Xianzhe_2024,Chen_Dissipationless_2015,Sajid_LBFO_2024,Meisenheimer_MagnonAnisotropy_2024}; however, the microscopic mechanism of the magnon-mediated spin transport remains unknown. This is in contrast to ferro(ferri)magnets, where the spin carried by magnons is simply controlled by the magnetic field, and the microscopic origins of spin transport are well described \cite{Xiao_MagnonSSEtheory_2010,Rezende_SSEtheory_2018,Cornelissen_YIGmagnon_2015}. 

Despite the complexity of the magnetic structure in BFO \cite{Gross_realSpaceImagin_2017,Meisenheimer_PersistentAnisotropy_2024}, a simple phenomenological approach to magnon-mediated spin transport in a generalized magnetic texture would offer insight into the physical mechanisms of electric-field-controlled magnon transport. By applying mirror and time-reversal operations on the magnon propagation, it is possible to predict the behaviour of the spin current based on the transformations of the magnetic texture under these operations. We apply this to thermally excited magnon-spin transport in three model BFO samples with different spin-cycloid configurations, where the behavior of the polarization and the magnetic texture under different electric fields has been mapped. We find that the model's predictions, based on the symmetries associated with the samples' magnetic textures, match the measured magnon-transport data. These results show that this model can be a powerful tool to guide further studies into the microscopic origin of magnon-mediated spin transport, as well as predict the qualities of magnon transport in new multiferroic systems.
\begin{figure}[t!]
\centering
\includegraphics[width=0.48\textwidth]{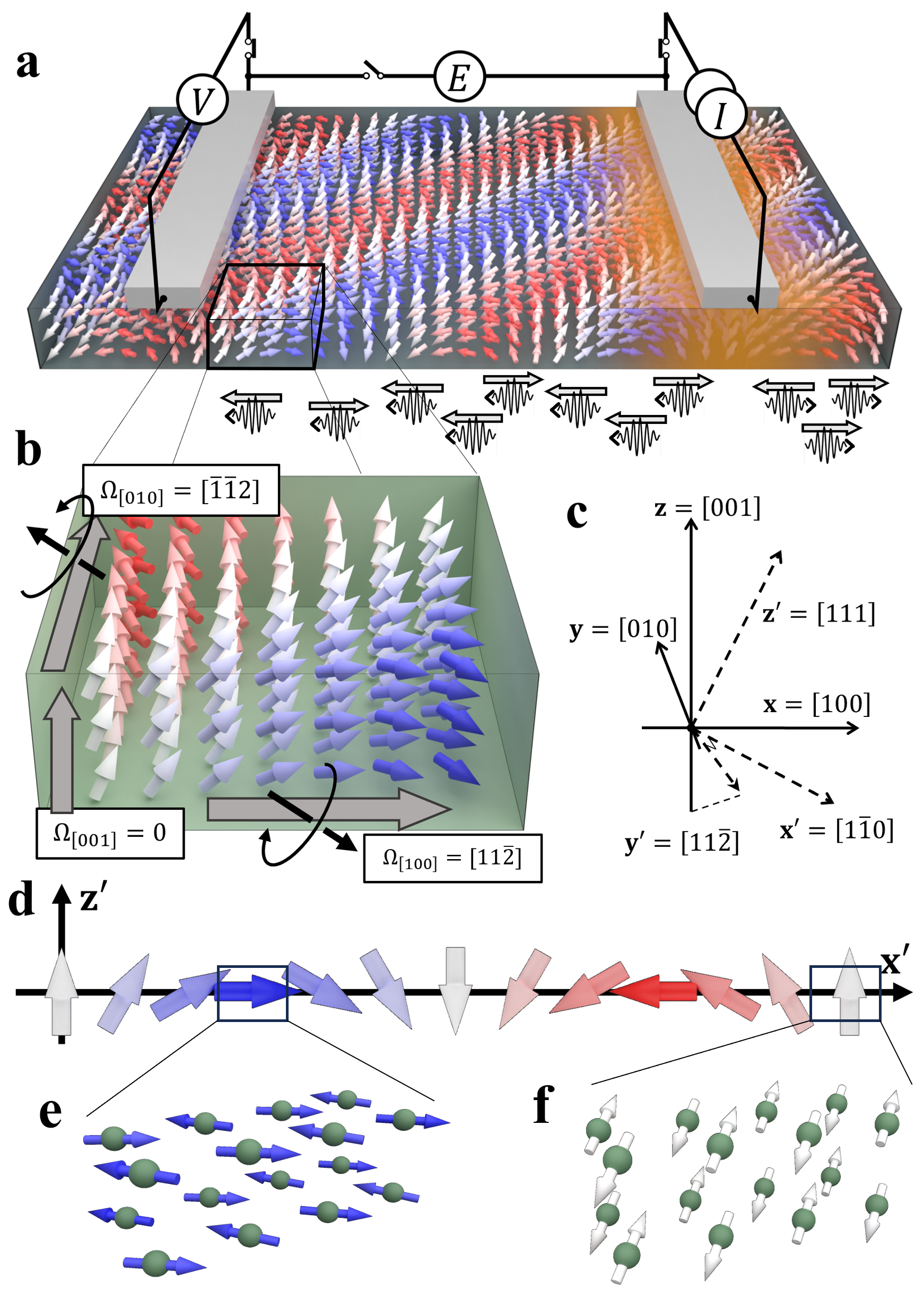}
\caption{\textbf{(a)} Magnons (represented by wavepackets here) with $\langle S_x\rangle \sim\pm\vu{x}$, as notated by the block grey arrows, diffuse along the temperature gradient through the BFO/BLFO and impart their spin to the detector, generating an ISHE voltage. \textbf{(b)} The colored arrows represent the local N\'eel vector in the BFO. The polarization is along $\vb{z}' = [111]$ and the cycloid propagation direction is $\vb{x}'=[1\bar{1}0]$. The red/blue coloring gives the net moment due to canting, as would be detected in NV microscopy. The vector $\vb*{\Omega}_{\vb{b}}$ represents the axis around which the N\'eel vector rotates along the direction $\vb{b}$.  Looking at the change of the N\'eel vector along the $\vb{b}=[100]$, $[010]$, or $[001]$ directions, the N\'eel vector rotates clockwise, counterclockwise, or not at all around $[\bar{1}\bar{1}2]$, so $\vb*{\Omega}_{[100],[010],[001]}=[11\bar{2}],[\bar{1}\bar{1}2],0$.  \textbf{(c)} The device axes has $\vb{x}$ parallel to the applied field and temperature gradient, $\vb{y}$ parallel to the Pt wires, and $\vb{z}$ as the film normal. Cycloid (primed) coordinates have $\vb{z}'$ parallel to the BFO polarization, $\vb{x}'$ parallel to the cycloid propagation axis, and $\vb{y}'=\vb{z}'\times\vb{x}'$ normal to the plane of the cycloid, which is shown in \textbf{d}.  \textbf{(e)} The atom to atom variation in magnetic moment from the spin cycloid drawn to scale, including the spin density wave (not drawn to scale). We note that in \textbf{a}, \textbf{b}, and \textbf{d}, the arrows represent the average local N\'eel vector, however in \textbf{e} and \textbf{f} they represent the atomic magnetic dipoles.}
\label{Fig:schem}
\end{figure}

Thin films of BFO and its derivative Bi$_{0.85}$La$_{0.15}$FeO$_3$ (BLFO) were deposited using pulsed-laser deposition and molecular-beam epitaxy (Supplemental Note 1 \cite{SM}), and platinum (Pt) was sputtered for voltage detection of spin currents arising from non-local magnon excitations \cite{Cornelissen_YIGmagnon_2015,Ross_SSEinHematite_2021,das2022anisotropic} (Fig. \ref{Fig:schem}a). An electric field is applied between the two wires to set the polarization state of the BFO (BLFO) while disconnecting the source current and voltage detector. Then, the electric field is turned off and the current source is activated across one wire, while the lock-in voltage detector is across the other. Here, voltage measurements reference the second harmonic of the low frequency (7 Hz) source current \cite{Cornelissen_YIGmagnon_2015,Ross_SSEinHematite_2021} to select magnons generated from the spin-Seebeck effect (SSE). This is repeated (automatically through a Keithley switchbox) over a range of electric fields to extract the inverse spin-Hall effect (ISHE) voltage ($V_{\ISHE}$) as a function of electric field, and then the identities of the source and detector wires are switched to extract the data from a thermal gradient applied in the opposite direction (Fig. \ref{Fig:tables_data} and Supplemental Note 2 \cite{SM}). 

To tie the observed magnon signal to the microscopic magnetic structure, we begin with a phenomenological model for magnon transport in a generic multiferroic. By extending the formalism from prior thermal magnon works \cite{Rezende_SSEtheory_2018,Adachi_SSEtheory_2013,Rezende_AFMmagnons_2019}, we write the spin-current density $\vb{j}_s(\vb{r})$ (traveling in the $z$-direction) from magnon modes indexed by $\mu$:
\begin{equation}\label{J_s}
    \vb{j}_s(\vb{r}) \approx \sum_\mu\langle \vb{S}\rangle_\mu \int d^3k\,\, v_{\mu k,z}\left(n_{\mu k}-n_{\mu k}^0\right).
\end{equation}
The vector part of $\vb{j}_s(\vb{r})$ denotes the spin polarization direction, $\langle \vb{S}\rangle_\mu$ is the expected value of spin carried by magnon mode $\mu$ \cite{NOTE1}, $v_{\mu k,z}$ is the $z$-component of the group velocity $\vb{v}_{\mu k} = \vu{k}\pdv{\omega_k}{k}$, $n_{\mu k}(\vb{r})$ is the total number of magnons and $n_{\mu k}-n_{\mu k}^0$ is the number of magnons in excess of equilibrium at any given location $\vb{r}$, in mode $\mu$ with wavevector $k$. Further details can be found in Supplemental Note 3 \cite{SM}. These non-equilibrium magnons, represented pictorially (Fig. \ref{Fig:schem}a), are responsible for the finite spin-current output. This spin-current density is integrated over the detector surface $S$ to get the total spin current $\vb{I}_s$ entering the detector wire:
\begin{equation}\label{introduce_eta}
    \vb{I}_s = \int_S d^2r \vb{j}_s(\vb{r})\approx\sum_\mu\langle\vb{S}\rangle_\mu\eta_\mu
\end{equation}
where
\begin{equation}\label{eta_def}
    \eta_\mu = \int_S d^2r d^3k\,\,v_{\mu k,z}\left(n_{\mu k}-n_{\mu k}^0\right).
\end{equation}
Here, we have introduced $\eta_\mu$ as the extent to which the magnon mode $\mu$ contributes its spin $\langle \vb{S}\rangle_\mu$ to the detector wire. As a phenomenological function, $\eta_\mu$ is dependent on a) the underlying effective Hamiltonian and the magnetic ground state $\Psi$, b) the direction of the magnon diffusion $\vu{q}$, and c) the device geometry, which is effectively constant throughout all of the studies.

\begin{figure}[t!]
\centering
\includegraphics[width=0.45\textwidth]{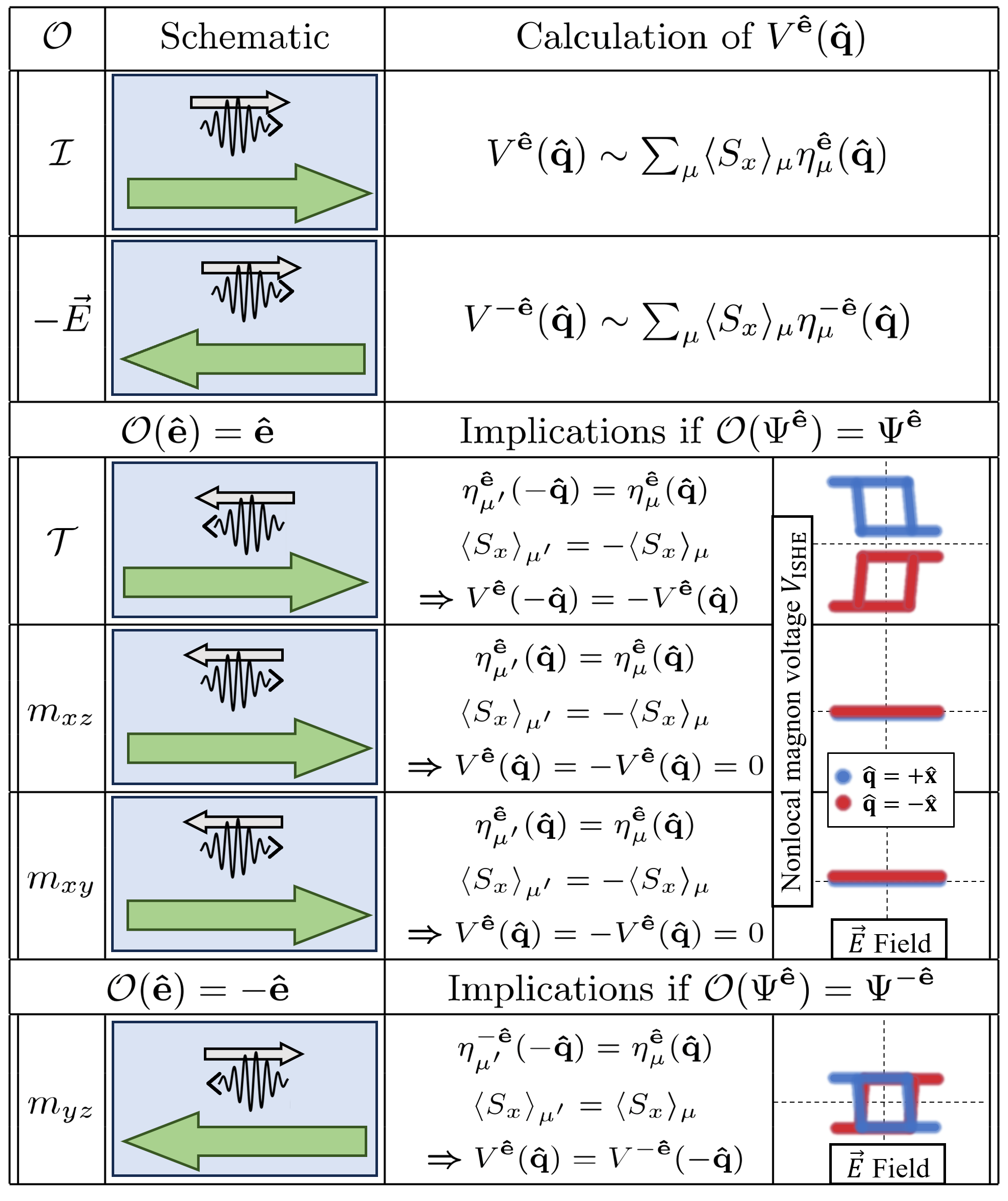}
\caption{\textbf{Symmetry operations} on a magnon mode $\mu$ with spin $\langle \vb{S}\rangle_\mu \parallel \vu{x}$ diffusing in the $\vu{x}$ direction, through a multiferroic poled along $\vu{x}$. The black arrow attached to the wavepacket denotes the magnon diffusion direction, the grey arrow denotes the magnon spin direction, and the large green arrow denotes poling direction. The rows show, from top to bottom, the identity, electric field poling, time reversal, and mirror operations over the $xz$, $xy$, and $yz$ planes. If the material (blue background) is invariant under the operations, we would expect each of the mathematical relations and their reflections on the measured magnon signal (as shown in the implications column) to hold true. We note that the $V^{\vu{e}}(\vu{q})$ values in the hypothetical magnon signals shown here are arbitrary.}
\label{Fig:symm}
\end{figure}

\begin{figure}[t!]
\centering
\includegraphics[width=0.45\textwidth]{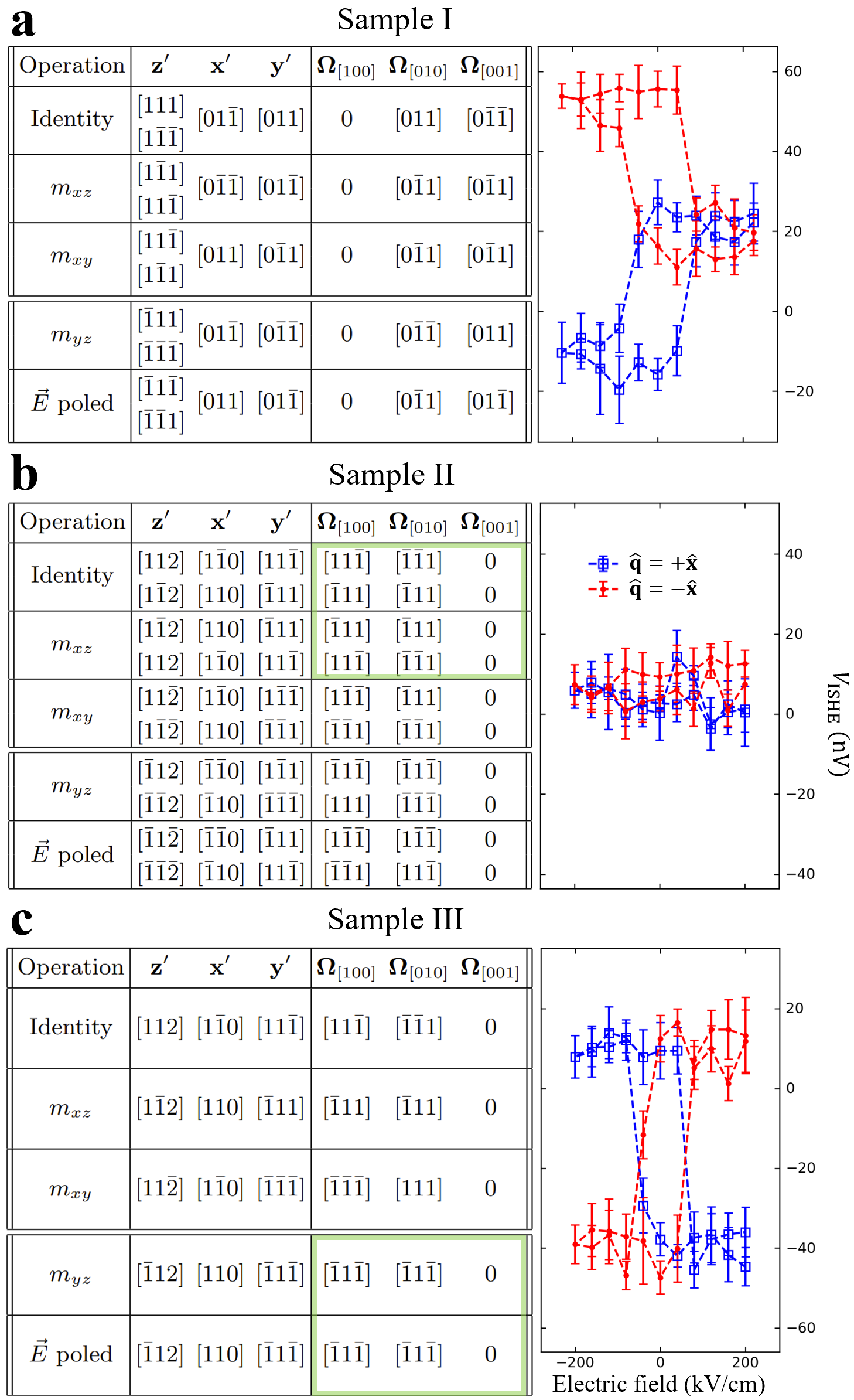}
\caption{\textbf{Calculation of} $\vb*{\Omega}_{\vb{b}}$\textbf{.} The polarization directions $\vb{z}'$ and cycloid propagation directions $\vb{x}'$, as well as the effects of $\Vec{E}$ poling, are taken from prior PFM and NV work \cite{Meisenheimer_MagnonAnisotropy_2024,Meisenheimer_PersistentAnisotropy_2024} on these samples, and Eq. \ref{BigOmega} is used to calculate $\vb*{\Omega}_{\vb{b}}$ for the three different samples. The $V_{\ISHE}$ hysteresis as a function of poling field for each sample is shown to the right for $\vu{q}=\pm\vu{x}$. All data is consistent with time reversal invariance, and sample II (III) shows $m_{xz}$ $(m_{yz})$ invariance in all three $\vb*{\Omega}_{\vb{b}}$ (and therefore invariance in $\vb*{\Omega}_{\vb{b}}$ for any $\vb{b}$, see Supplemental Note 6 \cite{SM}), as highlighted by the green boxes, and the implications from these invariances are reflected in the data.}
\label{Fig:tables_data}
\end{figure}

In a multiferroic, an electric field along $\vu{e}$ can be used to switch between different ferroelectric (polarization) states, which correspond to different magnetic ground states $\Psi^{\vu{e}}$. Furthermore, the direction of the thermal gradient ($i.e.$, the direction of magnon diffusion $\vu{q}$) can also be changed. In our experiment, we can alternate between $\vu{q}=\pm\vu{x}$ by switching the identity of the source wire and detector wire, thereby switching the direction of the temperature gradient. We switch the ferroelectric state by poling with positive or negative voltage across the detector and source wires, giving $\vu{e}=\pm\vu{x}$, where $\vu{e}$ is the direction of the poling field above the critical field (Supplemental Note 2 \cite{SM}). Thus, we write $\eta_\mu^{\vu{e}}(\vu{q})$ for the magnetic ground state $\Psi^{\vu{e}}$ as a function of $\vu{q}$. The nonlocal voltage, $V=R_{\textrm{Pt}}\theta_{\textrm{Pt}}(\vb{I}_s\cdot\vu{x})$ is then also a function of $\Psi^{\vu{e}}$ and $\vu{q}$ ($\mathcal{I}$ row of Fig. \ref{Fig:symm}). $R_{\textrm{Pt}}$ and $\theta_{\textrm{Pt}}$ are the resistance and the spin-Hall angle of Pt. We are only interested in the $\vb{x}$ component of $\vb{I}_s$ because the ISHE current is a cross product between the spin-current direction, $\vu{z}$, and the spin-current polarization, $\vb{I}_s$. While the magnitude of the nonlocal voltage $V$ will also depend on the absolute temperature and the magnitude of the temperature gradient, these variables are fixed in our experiments in order to focus on the symmetry-based phenomenological model.

Next, we consider symmetry operations on the ground state, experimental configuration, and magnon dynamics to impose constraints on the four values $V_{\ISHE} \equiv V^{\vu{e}}(\vu{q})$ (Fig. \ref{Fig:symm}) for $\vu{e},\vu{q}=\pm\vu{x}$. First, we consider the time reversal operation $\mathcal{T}$. A magnetic ground state with unpaired spins will break time-reversal symmetry, however, it is possible that the action of $\mathcal{T}$ on a magnetic texture is equivalent to a translation. For a translation in such a periodic magnetic texture with no global net magnetization, the magnon dynamics ($i.e.$, diffusion, spin transport) integrated over an area much larger than the periodicity of the texture will be invariant under the translation, and thus will also be invariant under $\mathcal{T}$.

For any thermal magnon mode $\mu$ in such a magnetic texture, the action of $\mathcal{T}$ will transform the mode $\mu$ into mode $\mu'$, with spin and diffusion reversed, as shown schematically ($\mathcal{T}$ row of Fig. \ref{Fig:symm}). Due to the invariance of the magnetic texture under $\mathcal{T}$, however, the dynamics encapsulated by $\eta$ will be the same for both modes. Summing over all modes $\mu'$ to get a nonlocal voltage, we find that $V^{\vu{e}}(\vu{q})=-V^{\vu{e}}(-\vu{q})$, which can be seen by combining the first two equations in the $\mathcal{T}$ row with the equation in the $\mathcal{I}$ row of Fig. \ref{Fig:symm}. When the two-hysteresis measurements of $V_{\ISHE}$ are made for such a multiferroic texture and the four voltage measurements are extracted, the above condition causes the polarity of the hysteresis to reverse, and the $\Delta V_{\ISHE}$ of the hysteresis to stay the same in switching from a $+\vu{q}$ measurement to a $-\vu{q}$ measurement as depicted in the $\mathcal{T}$ row of Fig. \ref{Fig:symm}.

\mycomment{
\begin{equation}\label{Time_spin}
\langle\vb{S}\rangle_\mu\to\langle\vb{S}\rangle_{\mu'}=-\langle\vb{S}\rangle_\mu.
\end{equation}}

\mycomment{
\begin{equation}\label{Time_eta}
\eta_{\mu}(\vu{q},\vu{p})=\eta_{\mu'}(-\vu{q},\vu{p}),
\end{equation}}

\mycomment{
\begin{equation}\label{Time_I_s}
\begin{split}
    \vb{I}_s(\vu{q},\vu{p})&=-\vb{I}_s(-\vu{q},\vu{p})\\
    \Rightarrow V_{\ISHE}(\vu{q},\vu{p})&=-V_{\ISHE}(-\vu{q},\vu{p})
\end{split}
\end{equation}}

Any symmetry operation $\mathcal{O}$ can be analyzed in this way to find implications in the magnetic texture on the $V_{\ISHE}$ measurements. This process is done for three mirror operations $m_{xz}$, $m_{xy}$, and $m_{yz}$ (Fig. \ref{Fig:symm}) over the $xz$, $xy$, and $yz$ planes relative to the device geometry (see the unprimed coordinate system in Fig. \ref{Fig:schem}c). We note that upon applying $m_{yz}$, the poling direction is flipped, so the magnetic texture $m_{yz}(\Psi^{\vu{e}})$ must be compared to the oppositely poled multiferroic texture $\Psi^{-\vu{e}}$, as indicated by the table (Fig. \ref{Fig:symm}). See Supplemental Figure 4  \cite{SM} for visualization of these symmetry operations on the spin cycloid.

To test this model, we choose a set of three samples with different magnetic textures (Supplemental Note 4 \cite{SM}). Sample I is a 50-nm-thick BFO film grown on a TbScO$_3$ (110) substrate, with wires patterned parallel to the 109$^\circ$ ferroelectric stripe domains \cite{Meisenheimer_MagnonAnisotropy_2024}. Samples II and III are 45-nm-thick BLFO films grown on DyScO$_3$ (110) substrates, with wires patterned parallel to [100]$_{\textrm{pc}}$  and [010]$_{\textrm{pc}}$ (pc: pseudocubic), respectively \cite{Sajid_LBFO_2024}. All subsequent vectors are written in pseudocubic notation. Samples I and III have one variant of spin cycloid within the poled area, but sample II has two variants, with propagation vectors as noted (Fig. \ref{Fig:tables_data}). The spin cycloid ground state of BFO, with average N\'eel vectors sketched in Fig. \ref{Fig:schem}d and with Fe magnetic moments sketched in Fig. \ref{Fig:schem}e,f, is given by Fishman $et$ $al.$ \cite{Fishman_BFOhamiltonian_2013}. Our first observation is that for all samples, $\mathcal{T}$, which flips each spin, is equivalent to a translation by half a period of the cycloid (Supplemental Note 5 \cite{SM}). Since the magnetic texture is thus invariant under $\mathcal{T}$, we expect the $V_{\ISHE}$ data to exhibit the corresponding implications as shown (Fig. \ref{Fig:symm}). 

Measurements of $V_{\ISHE}$ for the three samples are provided (Fig. \ref{Fig:tables_data}), and it can be seen that the polarity of the hysteresis is reversed while the magnitude of $\Delta V_{\ISHE}$ remains the same upon switching $\vu{q}$, as expected from the $\mathcal{T}$ invariance. While each sample has a 5-20 nV offset, we surmise that this could be from a gradient in the $\vb{z}$ direction and the resulting spin transport \cite{shan2017criteria}.

To analyze the mirror-symmetry operations, we take a closer look at the cycloidal texture in each domain. The precise determination of the polarization direction $\vb{z}'$ and the cycloid-propagation direction $\vb{x}'$ are discussed in Supplemental Note 4 \cite{SM}. The rotation of the N\'eel vector around the cycloid plane normal $\vb{y}'$ (Fig. \ref{Fig:schem}c) changes under the mirror operations (Fig. \ref{Fig:tables_data} and Supplemental Figure 4 \cite{SM}). 

Since $\vb{y}'$ depends on the choice of $\vb{x}'$, we define $\vb*{\Omega}_{\vb{b}}$ as the rotation of the Néel vector as measured along $\vb{b}$ (Fig. \ref{Fig:schem}b):
\begin{equation}\label{BigOmega}
    \vb*{\Omega}_{\vb{b}} = (\vb{b}\cdot\vb{x}')\vb{y}'.
\end{equation}
It is clear that this observable does not depend on the sign of $\vb{x}'$ chosen. Figure \ref{Fig:tables_data} presents calculations for the three different BFO samples of $\vb*{\Omega}_{\vb{b}}$ with $\vb{b}=[100],$ $[010]$, and $[001]$, in the two different experimental configurations ($\vu{e}=+\vu{x}$, Identity, or $\vu{e}=-\vu{x}$, $\Vec{E}$ poled) and under the three different mirror operations (acting on $\Psi^{\vu{e}}$ for $\vu{e}=+\vu{x}$). Although only the directions are recorded here, the results hold if the magnitudes are included. The corresponding $V_{\ISHE}$ data is included to the right for comparison. Sample I has $\sim20$-nm-wide stripy ferroelectric domains with polarizations $\vb{z}_1'$ and $\vb{z}_2'$ at a 109$^\circ$ angle that lead to a single variant spin cycloid with a propagation direction $\vb{x}'$ which is perpendicular to both $\vb{z}_1'$ and $\vb{z}_2'$ and a cycloidal plane given by $\vu{y}' = \tfrac{1}{2}(\vb{z}'_1+\vb{z}'_2)\times\vu{x}'$ \cite{Meisenheimer_MagnonAnisotropy_2024}. We find that under the action of any mirror symmetry, $\vb*{\Omega}_{\vb{b}}$ changes, and so we expect no further than the $\mathcal{T}$ implications in the signal. The data reflect that symmetry, and show only the signatures of $\mathcal{T}$ invariance.

Sample II has two types of larger ferroelectric domains, each with their own spin cycloid as given by the $\vb{z}'$, $\vb{x}'$, and $\vb{y}'$ in Fig. \ref{Fig:tables_data}b \cite{Sajid_LBFO_2024}. The $m_{xz}$ operation effectively maps the cycloids, quantified by $\vb*{\Omega}_{\vb{b}}$, in each domain onto each other, leaving the global magnetic structure invariant up to a domain exchange. Since the population of both domains is the same, a domain exchange leaves the magnon signal invariant, so the sample is effectively invariant under $m_{xz}$. The implication, as given by Fig. \ref{Fig:symm}, is that $V^{\vu{e}}(\vu{q})=-V^{\vu{e}}(\vu{q})=0$, and aside from the offset from the $\vb{z}$ gradient, the signal is uniformly zero as expected, despite a robust ferroelectric hysteresis (Supplemental Note 2 \cite{SM}). This is an important result: the phenomenological model can be used to predict a lack of $V_{\ISHE}$ switching based on the symmetry of the magnetic texture, even without a microscopic understanding of the physics of magnon-spin transport. 

Sample III has a single ferroelectric domain with one variant of spin cycloid \cite{Sajid_LBFO_2024}. Notably for this sample, the effect of the $m_{yz}$ operation on $\vb*{\Omega}_{\vb{b}}$ is identical to the effect of opposite field poling; $m_{yz}(\Psi^{\vu{e}})=\Psi^{-\vu{e}}$. The implications (Fig. \ref{Fig:symm}) are indeed reflected in the data. The main difference between sample II and III is in the global magnetic texture, and our model identifies that the reduced symmetry of the magnetic texture in sample III allows for nonzero $V_{\ISHE}$. 

We note that in comparing predictions to experimental data, the model is limited by a) the ability of the spin-current absorbing contacts to average over the periodicity of the texture (as previously discussed) and b) any magnetic anomalies created by symmetry-breaking defects. Such defects will add signals that do not obey the implications of the symmetries broken. Despite this, the model still guides the overall understanding and predictions of the physical origin of magnon spin transport in these complicated magnetic textures. For example, we apply these same ideas to predict the detection of magnons created by the spin-Hall effect in the source wire (Supplemental note 7 \cite{SM}).

In conclusion, we have developed a phenomenological model for magnon-mediated spin transport in multiferroics, which summarizes the dynamics of a magnon mode $\mu$ with a phenomenological function $\eta_\mu$ of the experimental configuration. We have shown how this simple model, paired with an analysis of the magnetic texture based on symmetry operations, helps us to explain the behavior of magnon-mediated spin currents. We find that the model's predictions match well with the experimental data for second harmonic non-local magnon transport in BFO/Pt based systems. The approach can be extended generally to electric-field-controlled magnon propagation in all multiferroics, and will serve as an important tool for understanding spin currents in future magnon transport studies.

This work was primarily supported by the U.S. Department of Energy, Office of Science, Office of Basic Energy Sciences, Materials Sciences and Engineering Division under Contract No. DE-AC02-05CH11231 within the Quantum Materials Program (No. KC2202) and U.S. Department of Energy, Office of Science, Office of Basic Energy Sciences, Materials Sciences and Engineering Division under Contract No. DE-AC02-05-CH11231 (Codesign of Ultra-Low-Voltage Beyond CMOS Microelectronics (MicroelecLBLRamesh)) for the development of materials for low-power microelectronics. L.W.M. and R.R. also acknowledge partial support from the Army/ARL as part of the Collaborative for Hierarchical Agile and Responsive Materials (CHARM) under cooperative agreement W911NF-19-2-0119 and the National Science Foundation under Grant DMR-2329111. Second Harmonic generation experiments performed at the Molecular Foundry were supported by the Office of Science, Office of Basic Energy Sciences, of the U.S. Department of Energy under Contract No. DE-AC02-05CH11231. S.K.K. was supported by the Brain Pool Plus Program through the National Research Foundation of Korea funded by the Ministry of Science and ICT (2020H1D3A2A03099291), National Research Foundation of Korea(NRF) grant funded by the Korea government(MSIT) (2021R1C1C1006273), and National Research Foundation of Korea funded by the Korea Government via the SRC Center for Quantum Coherence in Condensed Matter (RS-2023-00207732). J.\'I.G. is supported by the Luxembourg National Research Fund (FNR) through Grant C21/MS/15799044/FERRODYNAMICS.

\bibliographystyle{apsrev4-1}
\bibliography{aps.bib}

\begin{thebibliography}{35}
\bibitem [{NOT()}]{NOTE1}%
  \BibitemOpen
  \href@noop {} {}\bibinfo {note} {Here we assume that the spin carried by the magnon is independent of the wavevector $k$ of the magnon, which is not always the case: N. Okuma, Phys. Rev. Lett. \textbf{119}, 107205 (2017).}\BibitemShut {Stop}%
\end{thebibliography}

\end{document}